\documentstyle[12pt]{article}
\begin{document}
\baselineskip=21pt
%
\newcommand{\bc}{\begin{center}}
\newcommand{\ec}{\end{center}}
\newcommand{\be}{\begin{equation}}
\newcommand{\ee}{\end{equation}}
\newcommand{\bq}{\begin{eqnarray}}
\newcommand{\eq}{\end{eqnarray}}
%
\section{\bf Introdu\c c\~ao}
\vskip 5mm

Este trabalho foi motivado pelo de 
\rm Ach\'ucarro, \rm Gregory e \rm Kuijken, \cite{5Gregory},
onde se considera uma corda c\'osmica na gravita\c c\~ao de Einstein
penetrando um buraco negro de Schwarzschild. 

Conclui-se que os 
campos do vortex constituem um cabelo do buraco negro \cite{5Gregory}
e que tal n\~ao \'e nenhuma excep\c c\~ao \`a conjectura dos n\~ao cabelos
na Relatividade Geral. 

Recorde-se que a conjectura dos n\~ao cabelos
afirma que a \'unica informa\c c\~ao de longo alcance
proveniente de um buraco negro \'e a proveniente apenas
da sua massa, carga e momento angular \cite{5Chrusciel}.
Assim por exemplo bons numeros quanticos para uma estrela de neutr\~oes
tais como o numero lept\'onico e barionico n\~ao o s\~ao para buracos negros.

\eject
 
\section{\bf Cordas que penetram buracos negros}
 
\vskip 5mm

Usando o mesmo m\'etodo que em \cite{5Gregory} mostramos \cite{5Santos} que
nas proximidades do core de uma corda fina dilat\'onica
\cite{3GregorySantos}
imersa no espa\c co tempo de um buraco negro
de Schwarzschild as solu\c c\~oes dos campos do vortex s\~ao
do tipo de Nielsen-Olesen \cite{1NOlesen} e constituem
de facto um cabelo do buraco negro.

Assim nas proximidades do core da corda
com densidade linear de energia $\mu$ e formada a escala de
energia $\eta = \sqrt{2\epsilon}$, 
a m\'etrica \'e dada pela de um buraco negro de Schwarzschild
com um deficit c\'onico de $4 \pi \epsilon {\hat \mu}$ 
que nas coordenadas de Schwarzschild tambem se escreve
\be 
ds^2 = (1-\frac{2{\hat E}}{{\hat r}}) d{\hat t}^2
- (1-\frac{2{\hat E}}{{\hat r}})^{-1} d{\hat t}^2
- {\hat r}^2 d\theta^2
- {\hat r}^2 (1 - A)^2 \, e^{-2C} \, \sin^{2}\theta d\varphi^2
\ee
onde ${\hat t} = e^{\frac{C}{2}} t$, $etc$, com $C$ uma quantidade 
positiva relacionada com a press\~ao radial da corda, ${\cal P}_{R}$,
$C = \epsilon \int_0^R R {\cal P}_{R}$ sendo
$R = r\sin\theta$. 

Assim a massa inercial do 
buraco negro $E_I = {\hat E} ( 1 - 4 \epsilon {\hat \mu})$
e menor que a gravitacional ${\hat E}$ \cite{5Gregory}.

Generalizamos estes resultados a um
buraco negro dilat\'onico magneticamente
carregado \cite{5Santos}.

{\bf 1} {\it dma3cds@fourier.dur.ac.uk},
On leave from:
Departamento de F\'\i sica da Faculdade de Ci\^encias da Universidade do
Porto,
Rua do Campo Alegre 687, 4150-Porto, Portugal.
 
\end {document}